\documentclass[reprint,
onecolumn,
 amsmath,amssymb,
aps,
prfluids
]{revtex4-1}

\usepackage{graphicx}% Include figure files
\usepackage{dcolumn}% Align table columns on decimal point
\usepackage{bm}% bold math
\usepackage[normalem]{ulem}
\begin{document}

\preprint{APS/123-QED}

\title{Generation of shear flows and vortices in rotating anelastic convection}% Force line breaks with \\

\author{Laura K. Currie}
 \email{L.K.Currie@exeter.ac.uk}
\affiliation{%
Department of Mathematics and Computer Science, College of Engineering, Mathematics and Physical Sciences, University of Exeter, EX4 4QF, UK
}%


\author{Steven M. Tobias}
 \email{S.M.Tobias@leeds.ac.uk}
\affiliation{Department of Applied Mathematics, University of Leeds, LS2 9JT, UK
}%

%\affiliation{%


\date{\today}% It is always \today, today,

\begin{abstract}
We consider the effect of stratification on systematic, large-scale flows generated in anelastic convection. We present results from three-dimensional numerical simulations of convection in a rotating plane layer in which the angle between the axis of rotation and gravity is allowed to vary. This model is representative of different latitudes of a spherical body. We consider two distinct parameter regimes: (i) weakly rotating and (ii) rapidly rotating. In each case, we examine the effect of stratification on the flow structure and heat transport properties focussing on the difference between Boussinesq and anelastic convection. Furthermore, we show that regimes (i) and (ii) generate very different large-scale flows and we investigate the role stratification has in modifying these flows. The stratified flows possess a net helicity not present in the Boussinesq cases which we suggest, when combined with the self-generated shear flows, could be important for dynamo action.

%\begin{description}
%\end{description}
\end{abstract}

\maketitle

%\tableofcontents

%\section{\label{sec:level1}First-level heading:\protect\\ The line
%break was forced \lowercase{via} \textbackslash\textbackslash}

%%%%%%%%%%%%%%%%%%%%%%
%%%%%  INTRODUCTION
%%%%%%%%%%%%%%%%%%%%%%
\section{Introduction}\label{sec:intro}

Fluid turbulence is known to interact non-trivially with rotation to drive systematic flows. The presence of rotation can lead to the breaking of the parity symmetry of the turbulent system and also to the presence of a pseudo-scalar that naturally leads to the establishment of mean flows.
In geophysical and astrophysical fluids, systematic flows coexisting with turbulence on a vast range of spatial and temporal scales are often observed. Examples include the differential rotation of the solar interior \cite{Schouetal1998}, the systematic banded zonal jets and polar vortices visible at the surface of the gas giants \cite{Porco2003, Vasavada2005, Adrianietal2018} and the strong flows driven in Earth's atmosphere and oceans \citep{Vallis2006}.
An important question is to identify the role of correlations in the turbulence in driving the mean flows \textit{and} the back-reaction of the mean flows on the statistics of the turbulence that leads to the self-consistent saturation of such flows. 

In many, though certainly not all, cases of interest the turbulent flow is driven by buoyancy forces and arises as thermal or compositional convection. The convective turbulence naturally interacts with rotation to drive mean flows, which themselves play a role in mediating the heat transport. This interaction is complicated and can even lead to such complicated dynamics as `predator prey bursting' where the solution oscillates between strong mean flows and efficient convection \citep{BrummellHart1993,GroteBusse2001,RotvigJones2006}. 

There have been many previous studies of the interaction of convection with large-scale systematic flows, owing to its key role in the dynamics of planetary and stellar interiors. 
These studies have employed a variety of techniques including Direct Numerical Simulation (DNS) \citep{BrunToomre2002, Miesch2005}, turbulence closure models \citep{KitchatinovRuediger1995,Rempel2005} or Direct Statistical Simulation (DSS) \citep{Tobiasetal2011, TobiasMarston2013, Marstonetal2016}. 
Investigations have been performed in both spherical geometry \cite{Miesch2000,Elliottetal2000,Christensen2001,Christensen2002,BrunToomre2002,Browningetal2004,GastineWicht2012,Gastineetal2013, Yadavetal2015, Dietrichetal2017}  and in local Cartesian models \cite{HS1983,HS1986,HS1987,JulienKnobloch,SaitoIshioka2011,curtob:2016,CurrieTobias2019}. There is also a class of intermediate local models that captures some of the effects of the vortex stretching engendered by the spherical geometry by use of an annulus geometry \cite{Busse1970, Jonesetal2003, Tobiasetal2018}.

In this paper we study the interaction of convection and systematic flows in a local Cartesian model. Such a model lacks some of the geometrical effects of the global spherical models, but allows for a higher degree of turbulence and for more rapid rotation than the more computationally demanding spherical models. The previous studies investigating the interaction of convection with systematic flows have predominantly been carried out utilising the Boussinesq approximation \cite{Boussinesq1903,SpiegelVeronis}, which is a good approximation when the density variations are relatively unimportant. This is the case for some (though not all) planetary interiors; for example, in the Earth's core the density varies by only approximately $20\%$ from the inner core boundary to the core-mantle boundary \citep{anufrievetal2005}. This translates to a number of density scale heights, $N_{\rho}$, of 0.18. By contrast, Jupiter's deep interior has a density contrast charaterised by $N_{\rho}\sim 5$ \citep{Frenchetal2012,Duarteetal2018} and so the Boussinesq approximation should be deemed less appropriate to this case. Similarly, density variations can also play an important role in the interior of stars (the convection zone of the Sun encompasses approximately 14 density scale heights \citep{ChristensenDalsgaardetal1996, Parker2001, AndersBrown2017}).

Relevant Boussinesq models include asymptotic models of rapidly rotating convection in a plane layer \citep{JulienKnobloch}, DNS of convection in layers with tilted rotation vectors and no slip \citep{HS1983} or stress-free boundary conditions \citep{Novietal2019} and the interaction of convection with an imposed shear flow \citep{HS1987, SaitoIshioka2011}. Furthermore \citet{mythesis} examined the generation of mean flows by Reynolds stresses in Boussinesq convection both in the absence and in the presence of a thermal wind and highlighted the importance of  the fluid Prandtl number and the angle of the rotation vector from the vertical for determining the dynamics.

In this paper we focus on the important role of stratification in determining the form of the convection and the associated flows. \citet{curtob:2016} considered the effect of stratification on mean flow generation in two-dimensional anelastic convection in a rotating plane layer. Here we extend that investigation to three dimensions and probe a larger range of rotation rates. 
The anelastic approximation allows the filtering of sound waves, whilst still retaining the effects of density stratification and so gives a computationally efficient framework for studying the role of stratification and its interaction with rotation.
Although fully compressible local models with rotation have been studied in the past \cite{Brummelletal1996,Brummelletal1998,Chan2001}, the local dynamics of anelastic convection is comparatively poorly studied and understood. 
However, we note that the anelastic framework has previously been utilised to model penetrative convection \citep{RogersGlatzmaier2005}, and for non-rotating two-dimensional systems \citep{Rogersetal2003}. Moreover, \citet{VerhoevenStellmach2014} performed 2D anelastic simulations of rapidly rotating convection in the equatorial plane (gravity and rotation perpendicular). More recently, \citet{Kessaretal2019} examined the role of stratification in determining the length scales of turbulent convection using 3D simulations of non-rotating anelastic convection in a Cartesian layer.

The main focus of this paper then is in determining the role of stratification in modifying interactions between large-scale shear flows (or vortices) and convection driven turbulence.
The paper is organised as follows: in section \ref{sec:model}, we present the model and governing equations. In section \ref{sec:Results}, we consider the effects of stratification on convection in two distinct regimes: (i) weakly rotating and (ii) rapidly rotating. Finally, we investigate the effect of stratification on large-scale flows that are self-consistently generated by the convection and discuss the potential of such flows to act as dynamos.

%%%%%%%%%%%%%%%%%%%%%
%%%%%  Model Setup
%%%%%%%%%%%%%%%%%%%%%
\section{Model setup and equations}\label{sec:model}

\begin{figure}
    \centering
    \includegraphics[scale=0.7]{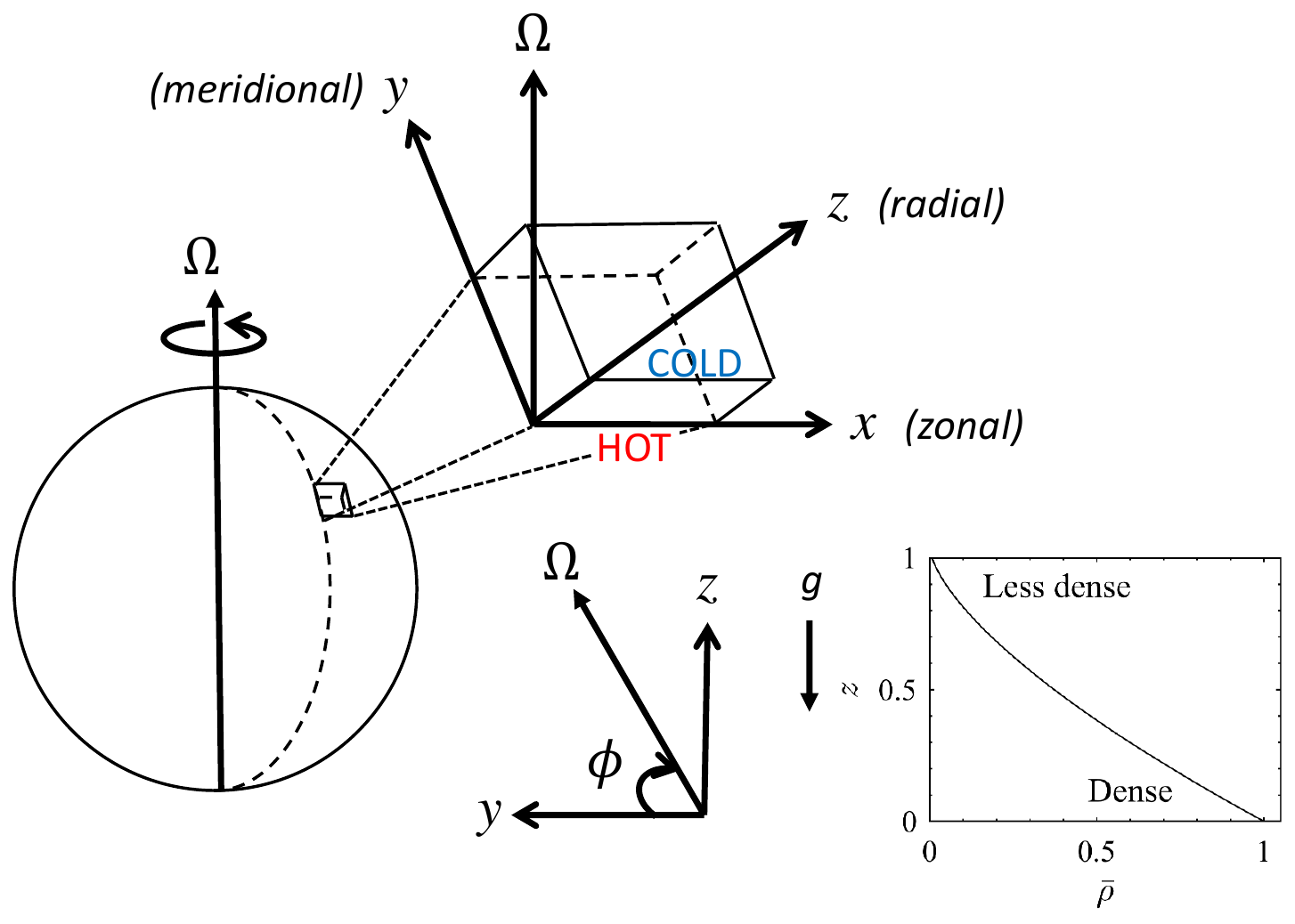}
    \caption{Geometry. The simulation domain is taken to be a Cartesian plane layer, located at a latitude $\phi$ on a spherical body. The rotation axis lies in the $y$-$z$ plane and is given (for a rotation rate $\Omega$) by $\bm{\Omega}=(0,\Omega\cos\phi,\Omega\sin\phi)$. The $z$-axis points upwards, the $x$-axis eastwards and the $y$-axis northwards. Gravity acts downwards. The axes on the right-hand-side show the background density stratification, $\bar\rho(z)$, for a strongly stratified system where the fluid is much denser at the bottom of the layer than at the top.}
    \label{fig:geometry}
\end{figure}

We consider a Cartesian plane layer of convecting fluid rotating about an axis that is oblique to gravity, which acts downwards. The rotation axis lies in the $y$-$z$ plane and is given by $\bm{\Omega}=(0,\Omega\cos\phi,\Omega\sin\phi)$, where $\Omega$ is the rotation rate and $\phi$ is the angle of the tilt of the rotation vector from the horizontal, so that the layer can be interpreted as being tangent to a sphere at a latitude $\phi$. We take the $z$-axis to point upwards, the $x$-axis eastwards and the $y$-axis northwards (see Figure \ref{fig:geometry}).
Note we choose the $z$-axis to point upwards for ease of comparison with Boussinesq models where the vertical axis increases upwards; this is in contrast to many compressible studies where it is taken to point downwards.

%%%%%%%%%%%%%

%%%%%% ANELASTIC EQUATIONS

%%%%%%%%%%%%%%

The anelastic equations are found by decomposing the density $\rho$, pressure $p$, and temperature $T$, of the fluid into an almost adiabatic reference state (denoted by an overbar) and a perturbation (denoted by a prime):
\begin{equation}
    \rho=\rho_0(\bar \rho + \epsilon \rho'), \mbox{\quad} p=p_0(\bar p + \epsilon p'), \mbox{\quad} T=T_0(\bar T + \epsilon T').
\end{equation}
The entropy is expanded as $s=const + \epsilon c_p(\bar s +s')$, where $c_p$ is the specific heat capacity at constant pressure. Here quantities with a subscript $0$ are reference values of the corresponding quantities taken at the bottom of the layer, $z=0$. $\epsilon$ is a measure of the departure of the reference state from adiabaticity, defined as follows:
\begin{equation}\label{epsilondef}
 \epsilon\equiv\frac{d}{H_0}\left(\frac{\partial \ln  \bar T}{\partial \ln \bar p}-\frac{\partial \ln \bar T}{\partial \ln 
\bar p}\bigg|_{\text{ad}}\right)=-\frac{d}{T_0}\left[\left(\frac{d \bar T}{dz}\right)_0+\frac{g}{c_p}\right]=-\frac{d}{c_p}\left(\frac{d \bar s}{dz}\right)_0\ll 1,
\end{equation}
where $d$ is the layer depth, $H=\frac{\bar p}{g\bar\rho}$ is the pressure scale height and $g$ is the acceleration due to gravity.
We assume an ideal gas so that $p=\mathcal{R}\rho T$ (where $\mathcal{R}$ is the universal gas constant); the entropy of the ideal gas is then given by $s=const + c_v\ln(\frac{p}{\rho^{\gamma}})$, where $c_v$ is the specific heat capacity at constant volume.
Note the constant in the definition of $s$ can be conveniently chosen so that $s=0$ on the upper boundary.

The anelastic governing equations (see for example \citep{Lantz1992, BragRob1995, LantzFan1999, MizerskiTobias2011}) for a fluid with velocity $\mathbf{u}=(u,v,w)$ can then be written as
\begin{align}\label{newmom}
\left[\frac{\partial \mathbf{u}}{\partial t}+(\mathbf{
u}\cdot\nabla)\mathbf{ u} \right]=-\nabla\left(\frac{
p}{\bar\rho}\right)+RaPrS\mathbf{\hat e_z}-Ta^{\frac{1}{2}}Pr\bm{
\Omega}\times\mathbf{
u}+\frac{Pr}{\bar\rho}\nabla\cdot{\bm\varsigma},
\end{align}
\begin{equation}\label{newcont}
 \nabla\cdot(\bar\rho{\mathbf{u}})=0,
\end{equation}
\begin{align}\label{newenergy}
\bar\rho \bar T\left[\frac{\partial{S}}{\partial  t}+(\mathbf{
u}\cdot\nabla)S\right]
=\nabla\cdot[\bar T\nabla S]-\frac{\theta}{\bar\rho
Ra}\frac{\bm{\varsigma}^{2}}{2},
\end{align}
where we have removed the primes from the perturbation quantities. In this formalism, we have assumed that the kinematic viscosity and the thermal conductivity do not vary with depth; this is discussed further below.
Note, because the reference state is close to adiabatic it is not necessary to introduce a separate perturbation to the entropy and so we instead solve for the total entropy $S=\bar s + s'$.
The full thermodynamic state can then be obtained from the anelastic versions of the equation of state and entropy definition respectively:
\begin{equation}
 \frac{p}{\bar p} = \frac{T}{\bar T}+ \frac{\rho}{\bar\rho},
\end{equation}
\begin{equation}\label{news}
S=\frac{1}{\gamma}\frac{p}{\bar p} - \frac{\rho}{\bar\rho}.
\end{equation}

The equations (\ref{newmom}) - (\ref{news}) are written in a dimensionless form, where we have scaled the dimensional equations using $d$ as the unit of length and the thermal diffusion time, $d^2/\kappa_0$, as the unit of time ($\kappa_0$ is the value of the thermal diffusivity at the bottom of the layer). 
$\bm\varsigma$ is the stress tensor defined by $\varsigma_{ij}= \bar\rho \left[ \frac{\partial u_i}{\partial x_j}+\frac{\partial u_j}{\partial x_i} - \frac{2}{3}(\nabla\cdot \mathbf u)\delta_{ij}\right]$ with $\bm\varsigma^2\equiv \bm\varsigma:\bm\varsigma =\varsigma_{ij}\varsigma_{ij}$. $\theta$ is the dimensionless temperature difference across the layer and $\gamma$ is the ratio of specific heats at constant pressure to constant volume. 
For convenience, we have introduced the dimensionless parameters:
\begin{equation}\label{eqn:dimlessparams}
 Ra=\frac{gd^3\epsilon}{\kappa_0\nu},
 \mbox{\quad }
 Ta=\frac{4\Omega^2d^4}{\nu^2}
\mbox{\quad and \quad}
     Pr=\frac{\nu}{\kappa_0},
    \end{equation}
commonly known as the Rayleigh, Taylor and Prandtl numbers respectively. Here $\nu$ is the kinematic viscosity which does not vary with depth in this model.

 %
%%%%%%%%%%%%%%%%%%%%%%%%% 
%%%%%%%%%%%%%%%%%%%%%%%%% 
%%%%% REFERENCE STATE
%%%%%%%%%%%%%%%%%%%%%%%%% 
%%%%%%%%%%%%%%%%%%%%%%%%% 

We consider a time-independent, polytropic reference state given by
\begin{align}
\bar T&=1+\theta z, \quad  \bar\rho= (1+\theta z)^m, \quad 
\bar p=-\frac{RaPr}{\theta(m+1)}(1+\theta z)^{m+1},\nonumber\\
\bar s &= \frac{m+1-\gamma m}{\gamma \epsilon}\ln(1+\theta z)+\text{const} \text{ \quad
with \quad} \frac{m+1-\gamma
m}{\gamma}=-\frac{\epsilon}{\theta}=\mathcal{O}(\epsilon),\label{eq:basicstateend}
\end{align}
where $m$ is the polytropic index, which we take to be 1.5 throughout this article, and $-1 < \theta \le 0$.
We note in this model the reference state is independent of time and so there is no adjustment by any mean that may be generated. The functional form of the background density stratification, $\bar\rho$, for a density contrast (between top and bottom of the layer) of approximately 150 is shown in Figure \ref{fig:geometry}.

\begin{figure}
    \centering
    \includegraphics{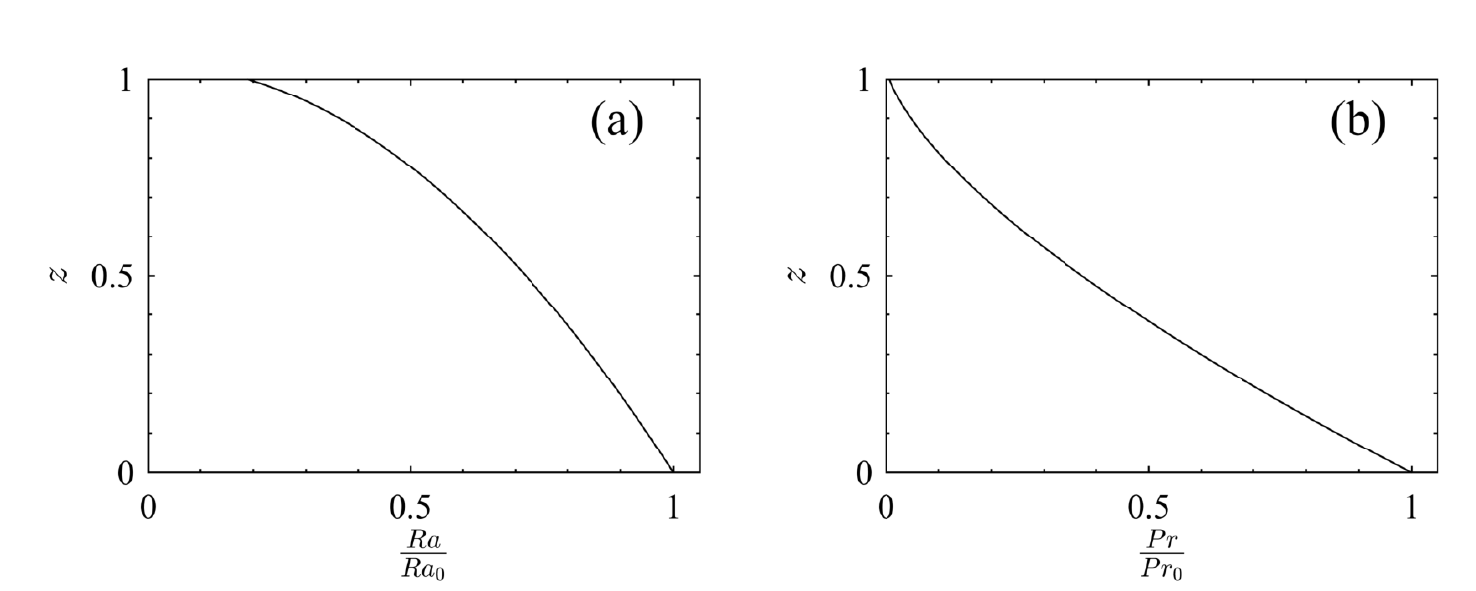}
    \caption{Depth dependent Rayleigh (a) and Prandtl (b) numbers as given by (\ref{eqn:depthdepRa}). They are shown for $\theta=-0.96433$ ($N_{\rho}=5$) and have been normalised by the value on the bottom boundary.}
    \label{fig:RaPrdepth}
\end{figure}

Note, within the anelastic approximation, for a stratified layer one cannot take both the dynamic viscosity $\mu$ and the kinematic viscosity $\nu$ of the fluid to be constant; at least one must be depth-dependent (since $\mu=\bar\rho \nu$). Similarly, one cannot take both the thermal conductivity $k$ and the thermal diffusivity $\kappa$ of the fluid to be constant and so there is some freedom over which parameters are kept constant across the layer depth and which vary. The results can depend on these choices (see, e.g., \citet{GlatzmaierGilman1981}).
In our formalism, we assume $\nu$ and $k=\bar\rho c_p\kappa$ (where we interpret $\kappa$ as the turbulent thermal diffusivity) to be constant (and therefore $\mu$ and $\kappa$ must vary with depth).
Whilst the overall Rayleigh and Prandtl numbers quoted in this article are defined at the bottom of the layer (as given in (\ref{eqn:dimlessparams})), we can also define analogous, depth-dependent dimensionless parameters $Ra(z)$ and $Pr(z)$ and these can vary significantly with depth at strong stratifications (note $Ta$ is constant across the depth). In particular, we have
\begin{equation}\label{eqn:depthdepRa}
    Ra(z)= Ra_0 (1+\theta z )^{m-1},\quad Pr(z)=Pr_0 (1+\theta z )^{m}.
\end{equation}
We note that, for no stratification, $\theta=0$, and $Ra$ and $Pr$ are constant across the layer.
Figure \ref{fig:RaPrdepth} shows the variation in $Ra(z)$ and $Pr(z)$ across the layer for a density contrast of approximately 150 between top and bottom of the layer ($\bar\rho$ for this case is shown in Figure \ref{fig:geometry}). It is clear that both $Ra$ and $Pr$ are significantly smaller at the top of the domain than at the bottom in strongly stratified cases. This will have significant consequences for the form of the convection, and will be discussed further in section \ref{subsec:solutionregimes}.

Anelastic formalisms also differ depending on whether entropy or temperature is diffused in the energy equation \cite{BragRob1995}.
We choose a model that takes the turbulent thermal conductivity to be much larger than the molecular conductivity and so equation (\ref{newenergy}) contains an entropy diffusion term but not a thermal diffusion term; \citet{BragRob1995} discuss models including both terms. In choosing to diffuse entropy, temperature can be eliminated as a variable from the formulation using a technique introduced by \citet{Lantz1992} and \citet{BragRob1995} (see also \citet{LantzFan1999, JonesKuzanyan2009}).

The equations (\ref{newmom})-(\ref{news}) along with the reference state (\ref{eq:basicstateend}) are similar to those given in case (1) of \citet{MizerskiTobias2011} but here the rotation vector has been generalised to include cases where the rotation vector and gravity are not aligned, i.e., $\mathbf{\Omega}=(0,\cos\phi, \sin\phi)$. \citet{curtob:2016} solved the 2.5D version of these equations, i.e.\ the version with all three velocity components, but with a dependence of all variables only on $y$ and $z$.
In this particular formalism, the anelastic equations (\ref{newmom})-(\ref{news}) reduce to the Boussinesq equations in the limit $\theta\rightarrow0$ and so $\theta$ can be thought of as a measure of the degree of compressibility.  However, we choose instead to use the number of density scale heights in the layer, $N_{\rho}$, as a measure of the stratification. $N_{\rho}$ is related to $\theta$ through the following: $N_{\rho}=\ln(1+\theta)^{-m}$; it can also be expressed in terms of the density contrast $\chi=\frac{\bar \rho|_{z=0}}{ \bar \rho|_{z=1}}$ via $N_{\rho}=\ln \chi$.

We solve the above anelastic equations subject to impenetrable, stress free and fixed entropy boundary conditions. In particular,
\begin{align}\label{eq:velBCs}
    w=\frac{\partial u}{\partial z}=\frac{\partial v}{\partial z}=0 \mbox{ on } z=0,1\\
    S=\Delta S \mbox{ on } z=0,\mbox{\quad} S=0 \mbox{ on } z=1,
\end{align}
where $\Delta S=\bar s|_{z=1}=-\frac{1}{\theta}\ln(1+\theta)$.

Throughout this paper we consider a domain of size $2\pi \times 2\pi \times 1$ (in dimensionless units) with periodic boundaries in the horizontal directions. We solve the system described by above, with a pseudospectral numerical code written in Dedalus \footnote{http://dedalus-project.org}\cite{Dedalus2019}. We use a Crank-Nicolson, Adams-Bashforth second order, semi-implicit time-stepping scheme. The time step is allowed to vary according to a standard CFL criterion. The spatial resolutions used are quoted in Table \ref{tab:table1} (note these values give the number of spectral modes after dealiasing; the number of grid point used can be found by multiplying this number by 3/2).

\subsection{Derived quantities}
In addition to the input dimensionless quantities given in (\ref{eqn:dimlessparams}), it is useful to define some diagnostic quantities that are derived from simulation output in order to help characterise the system.
The Rossby number is a measure of the strength of inertia relative to Coriolis forces and we calculate it as a function of depth from our simulations using the following definition:
\begin{equation}\label{eqn:Ro}
    Ro(z)=\left\langle \frac{\sqrt{|(\mathbf u \cdot \nabla)\mathbf u |^2}}{\sqrt{|\bm{\Omega} \times \mathbf u |^2}}\right\rangle_{x,y}
\end{equation}
where angle brackets $\langle \cdot \rangle_{x,y}$ denote an average over horizontal planes.
Since this quantity is unknown before the simulation has run, the so-called convective Rossby number, $Ro_c=\sqrt{Ra/(TaPr\sin\phi)}$ (for $\phi\neq 0$), is sometimes used as an a priori measure of the relative strength of inertia to Coriolis force. Note, for $N_{\rho}=0$ ($\theta=0$), $Ro_c$ is constant across the layer, but for $N_{\rho}\neq0$ ($\theta\neq0$), $Ro_c \propto (1+\theta z)^{-1/2}$ and so increases with $z$.
We will address the question of how good a proxy $Ro_c$ is for $Ro$ in our simulations is section \ref{subsec:solutionregimes}.

To quantify the effect of stratification on convective heat transport, we use the Nusselt number which we define as the ratio of the total heat flux to the conductive heat flux of the base state, i.e., 
\begin{equation}\label{eqn:Nusselt}
Nu=\frac{\frac{\partial S}{\partial z}}{\frac{\rm{d}\bar s}{\rm{d}z}}=-(1+\theta z)\frac{\partial S}{\partial z}.
\end{equation}
This gives $Nu$ at bottom of layer as $Nu|_{z=0}=-\frac{\partial S}{\partial z}|_{z=0}$ and $Nu$ at the top as $Nu|_{z=1}=-(1+\theta)\frac{\partial S}{\partial z}|_{z=1}$; we expect these to be equal in a steady state.

%%%%%%%%%%%%%%%%%%%%%
%%%%%%%%%%%%%%%%%%%%%
%%%%%  Results
%%%%%%%%%%%%%%%%%%%%%
%%%%%%%%%%%%%%%%%%%%%
\section{Results}\label{sec:Results}

We consider a series of numerical simulations using the setup described in section \ref{sec:model}. The onset of compressible convection in a local Cartesian geometry assuming the anelastic approximation varies depending on the precise form of the anelastic approximation  utilised \cite{KatoUnno1960, JonesRobertsGalloway1990, Calkins2014}. The linear problem for the particular formalism described in section \ref{sec:model} was studied in \citet{MizerskiTobias2011} for the case gravity and rotation aligned and extended in \citet{curtob:2016} to consider the case of tilted rotation. In this paper, we focus on the nonlinear regime and discuss the role of stratification in modifying the dynamics of convection in two cases: (i) a weakly rotating regime, where the large-scale flow takes the form of a vertical shear and (ii) a rapidly rotating regime, where it has the character of a large-scale vortex.
Within each regime, we consider tilted and untilted cases for a range of stratifications spanning $N_{\rho}=0$ (Boussinesq) to $N_{\rho}=5$. Within each subset of simulations, we fix $Ro_c$ at the bottom of the domain. 
We note that the Reynolds number of the two regimes is not comparable. The cases are selected to be at different convective Rossby numbers (at the bottom of the computational domain), and this leads to different dynamics in particular to the response of the convection to rotation. The dependence of Reynolds number on Rayleigh number is know to be a function of rotation rate \citep{Longetal2020}, and so any attempt to fix the Reynolds number across parameter regimes a priori is difficult.
Furthermore the Reynolds number will turn out to be a strong function of depth for stratified convection.
A summary of the input parameters used in our simulations is given in table \ref{tab:table1}. Each subset of simulations is given a name (A-D) for ease of reference and each simulation within each subset is given a number to reflect the value of $N_{\rho}$ (see first column of table \ref{tab:table1}).

\begin{table*}
\caption{\label{tab:table1}Table of simulation parameters. The first column gives each simulation a name: the letter being the set the simulation belongs to, and the number the value of $N_{\rho}$ in that simulation. We consider four sets of simulations A-D. A and B are from regime (i) - weakly rotating and C and D from regime (ii) - rapidly rotating. The second and third column give the value of $Ra$ and $Ta$ at the bottom boundary, respectively. The fourth column gives the latitude $\phi$. Sets A and C have the rotation axis aligned with gravity ($\phi=90^{\circ}$) while sets B and D have $\phi=45^{\circ}$. The fifth column gives the value of $N_{\rho}$; within each set we consider a range of $N_{\rho}$ from 0 to 5. The sixth and seventh columns give the value of $Ro_c$ on the bottom and top boundaries respectively. The final column gives the resolution $n_x \times n_y \times n_z$; here $n_i$ corresponds to the number of spectral modes used in the $i-$direction after 2/3 dealiasing has been applied. That is, 3/2 times $n_i$ is the number of grid points in physical spaces used in the $i-$direction. Note the only difference between simulations B0 and B0* and B5 and B5* was the state used to initialise each simulation. B0 and B5* were started from zero velocity and small entropy perturbations (and in particular, had no vertically integrated horizontal momentum initially), whereas B0* and B5 were started from states which possessed a non-zero vertically integrated horizontal momentum.}

\begin{tabular}{c c c c c c c c }
\hline\hline
 Name &  $Ra$ & $Ta$ & $\phi$  &  $N_{\rho}$ & $Ro_{c,bot}$ &  $Ro_{c,top}$ &  $n_x \times n_y \times n_z$     \\[0.3pt] \hline
    A0  &  $4\times 10^5$ & $4\times 10^4$  & 90 & 0 & 3.16 & 3.16 & $128 \times 128 \times 128$\\
    
    A1  &  $4\times 10^5$ & $4\times 10^4$  & 90 & 1 & 3.16 & 4.41 & $192 \times 192 \times 128$\\
    
    A2  &  $4\times 10^5$ & $4\times 10^4$  & 90 & 2 & 3.16 & 6.16 & $192 \times 192 \times 128$\\
    
    A3  &  $4\times 10^5$ & $4\times 10^4$  & 90 & 3 & 3.16 & 8.60 & $192 \times 192 \times 128$\\

    A4  &  $4\times 10^5$ & $4\times 10^4$  & 90 & 4 & 3.16 & 12.00 &$256 \times 256 \times 128$\\
  
    A5  &  $4\times 10^5$ & $4\times 10^4$  & 90 & 5 & 3.16 & 16.74 &$256 \times 256 \times 128$\\[5pt]
  
    B0/B0* &  $4\times 10^5$ & $4\times 10^4$  & 45 & 0 & 3.76 & 3.76 &$128 \times 128 \times 128$\\
    B1 &  $4\times 10^5$ & $4\times 10^4$  & 45 & 1 & 3.76 & 5.25 &$192 \times 192 \times 128$\\
    B2 &  $4\times 10^5$ & $4\times 10^4$  & 45 & 2 & 3.76 & 7.32 &$192 \times 192 \times 128$\\
    B3 &  $4\times 10^5$ & $4\times 10^4$  & 45 & 3 & 3.76 & 10.22 &$192 \times 192 \times 128$\\
    B4 &  $4\times 10^5$ & $4\times 10^4$  & 45 & 4 & 3.76 & 14.27 &$256 \times 256 \times 128$\\
    B5/B5* &  $4\times 10^5$ & $4\times 10^4$  & 45 & 5 & 3.76 & 19.91 &$256 \times 256 \times 128$\\[5pt]

    C0 & $1\times 10^7$ & $2\times 10^8$  & 90 & 0 & 0.22 & 0.22 &$384 \times 384 \times 128$\\
    C1 & $1\times 10^7$ & $2\times 10^8$  & 90 & 1 & 0.22& 0.31&$384 \times 384 \times 128$\\
    C2 & $1\times 10^7$ & $2\times 10^8$  & 90 & 2 & 0.22& 0.44&$512 \times 512 \times 128$\\
    C3 & $1\times 10^7$ & $2\times 10^8$  & 90 & 3 & 0.22& 0.61&$512 \times 512 \times 128$\\
    C4 & $1\times 10^7$ & $2\times 10^8$  & 90 & 4 & 0.22& 0.85&$512 \times 512 \times 192$\\
    C5 & $1\times 10^7$ & $2\times 10^8$  & 90 & 5 & 0.22& 1.18&$512 \times 512 \times 192$\\[5pt]

    D0 & $1\times 10^7$ & $2\times 10^8$  & 45 & 0 & 0.27 & 0.27 &$384 \times 384 \times 128$\\
    D1 & $1\times 10^7$ & $2\times 10^8$  & 45 & 1 & 0.27& 0.37&$384 \times 384 \times 128$\\
    D2 & $1\times 10^7$ & $2\times 10^8$  & 45 & 2 & 0.27& 0.52&$512 \times 512 \times 128$\\
    D3 & $1\times 10^7$ & $2\times 10^8$  & 45 & 3 & 0.27& 0.72&$512 \times 512 \times 128$\\
    D4 & $1\times 10^7$ & $2\times 10^8$  & 45 & 4 & 0.27& 1.01&$512 \times 512 \times 192$\\
    D5 & $1\times 10^7$ & $2\times 10^8$  & 45 & 5 & 0.27& 1.41&$512 \times 512 \times 192$\\ 
\hline\hline
\end{tabular}
\end{table*}

\subsection{Solution regimes}\label{subsec:solutionregimes}
\begin{figure}
    \centering
    \includegraphics{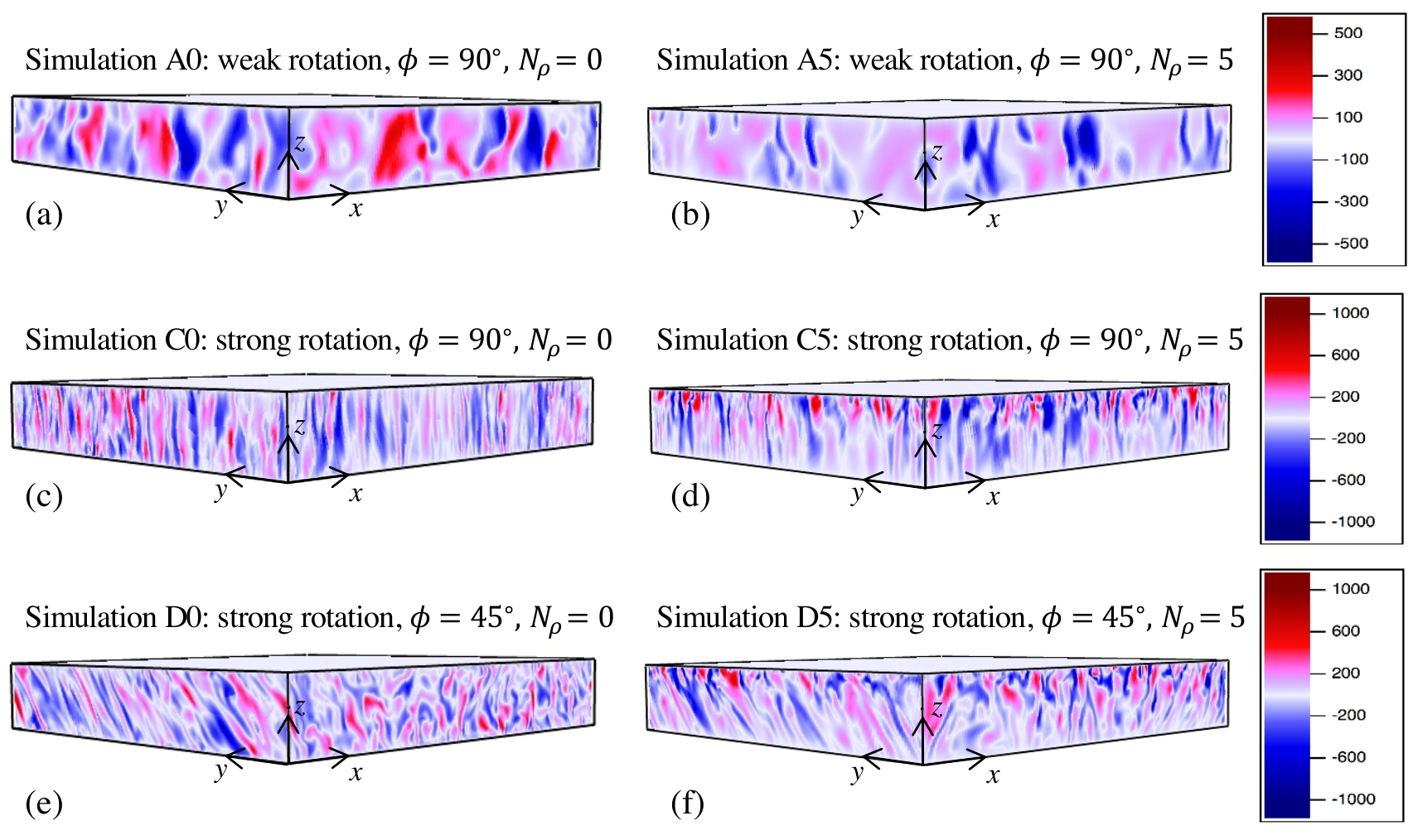}
    \caption{Snapshots of the vertical velocity, $w$, for 6 different simulations as defined in Table 1. (a) $Ro_{c,bot}=3.16$, $\phi=90^{\circ}$, $N_{\rho}=0$ (simulation A0); (b) $Ro_{c,bot}=3.16$, $\phi=90^{\circ}$, $N_{\rho}=5$ (simulation A5); (c) $Ro_{c,bot}=0.22$, $\phi=90^{\circ}$, $N_{\rho}=0$ (simulation C0); (d) $Ro_{c,bot}=0.22$, $\phi=90^{\circ}$, $N_{\rho}=5$ (simulation C5); (e) $Ro_{c,bot}=0.27$, $\phi=45^{\circ}$, $N_{\rho}=0$ (simulation D0); (f) $Ro_{c,bot}=0.27$, $\phi=45^{\circ}$, $N_{\rho}=5$ (simulation D5). Rotation causes the convection cells to align with the rotation axis but stratification disrupts the independence along this direction. }
    \label{fig:snapshots}
\end{figure}

Regimes (i) and (ii) lead to flows with very different morphologies, as perhaps expected. Typical snapshots of the vertical velocity from each regime are shown in Figure \ref{fig:snapshots}. The first two panels, (a) and (b), show the snapshots from a weakly rotating case for a Boussinesq simulation and for a simulation with $N_{\rho}=5$ respectively. In both cases the rotation is aligned vertically and the convection cells show alignment with the rotation axis but the horizontal length scale of the cells is broad. By contrast, Figure \ref{fig:snapshots} (c) and (d) show more rapidly rotating cases for a Boussinesq case and a case with $N_{\rho}=5$ respectively. Again, in both cases the rotation is aligned vertically. In these cases, thin convection cells are seen that align with the rotation axis as might be expected from the Taylor-Proudman theorem. We note that the Reynolds number for simulation set C are larger than in simulation set A and so the role of turbulence is stronger. However, a more significant effect is that the convective Rossby number, which gives a measure of the strength of advection relative to rotation is much smaller (at the bottom of the domain) in set C,  and this is reflected in the dynamics. In the Boussinesq case, these cells extend relatively uniformly across the height of the domain, whereas in the stratified case, the vertical velocity is larger at the top of the domain where the density is lower (this effect will be studied in more detail in section \ref{subsec:LSflows}). This effect of stratification is more evident in the rapidly rotating case than in the weakly rotating one, however even in the weakly rotating case, there is still an asymmetry in the layer.

In Figure \ref{fig:snapshots} (e) and (f), the parameters are the same as for (c) and (d), except for now the rotation vector is tilted to $45^{\circ}$. In this case, the convection cells are tilted in the $y-z$ plane to align approximately with the direction of the rotation vector; again a consequence of the strong rotation. The width of the convection cells in the bottom row are perhaps slightly larger than in the second row -- this is likely to be a consequence of the reduced vertical component of rotation. Again the effect of stratification is evident in the bottom right panel, where the flow velocity is stronger near the top of the layer.

The time taken for some of the simulations to reach a statistically-steady state from a small random initial perturbation can be long. In particular, in the rapidly rotating cases (regime (ii)), it can take several diffusion times for the kinetic energy to reach a saturated state. The likely reason for this is that in the rapidly rotating cases, a large-scale structure that evolves on the viscous timescale emerges from the convection. Similar structures have been observed in the work of \citep{Mantereetal2011,ChanMayr2013,Rubioetal2014,Favieretal2014,Guervillyetal2014,Novietal2019} and will be discussed in more detail in section \ref{subsec:LSflows}. Therefore, for computational convenience, we ran some fiducial cases starting from rest, and others starting from saturated states of previous runs. In contrast, the weakly rotating cases evolve to a statistically-steady state well within in a thermal/viscous diffusion time. 

\begin{figure}
    \centering
    \includegraphics[scale=1]{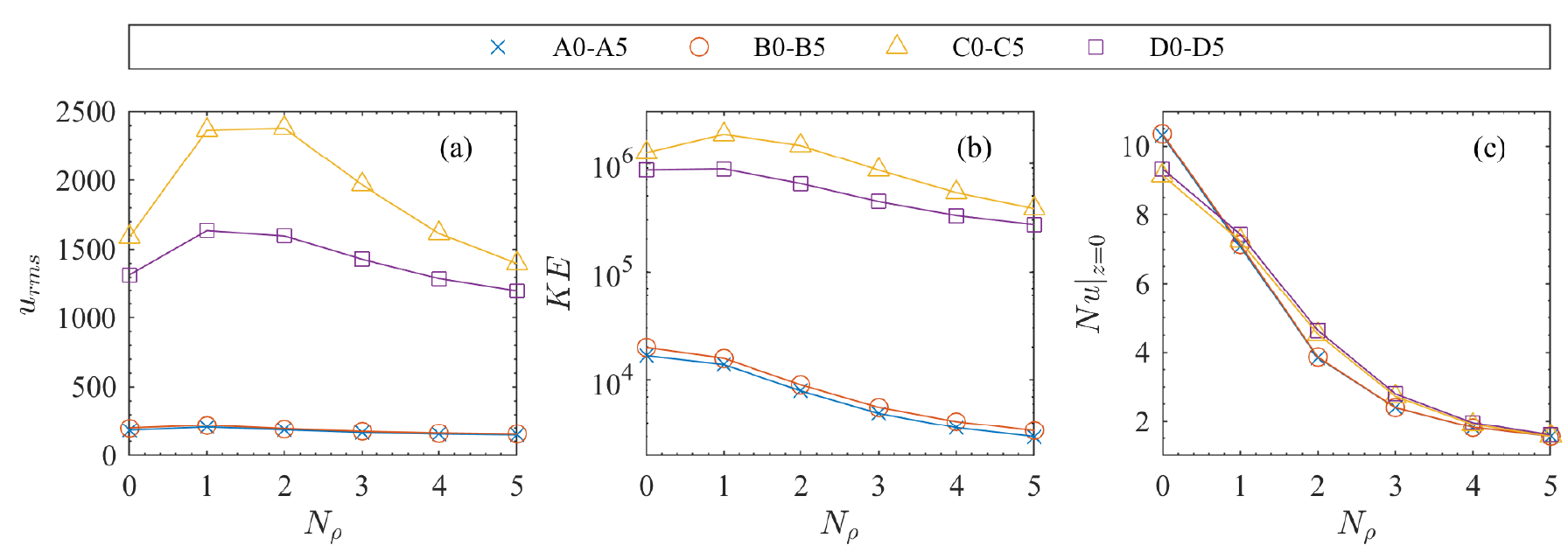}
    \caption{$u_{rms}$ (a), $KE$ (b) and $Nu$ (c) as a function of $N_{\rho}$ for simulations in each of the regimes: A (blue crosses), B (orange circles), C (yellow triangles) and D (purple squares).}
    \label{fig:energies}
\end{figure}

In figure \ref{fig:energies} we show time-averaged quantities where the temporal averages are taken over a time interval in which the solution is in a saturated state. We note that for each set of parameter values it may be possible that multiple states can be found (see e.g., \citep{Favieretal2019}) but we do not investigate this further here.
Figure \ref{fig:energies} (a) shows the time average of $u_{rms}$ for simulations in each of the regimes A-D. Here, $u_{rms}=\sqrt{\langle u^2+v^2+w^2\rangle}$, where the angle brackets denote a volume average. 
For all four regimes, the general trend is that as $N_{\rho}$ is increased from zero, $u_{rms}$ first increases, but then as $N_{\rho}$ is increased further, $u_{rms}$ decreases. This could be a result of competing effects: (i) as $N_{\rho}$ is increased, the average density is decreased which we would expect might lead to faster velocities but (ii) $Ra$ is not constant across the domain and is proportional to $(1+\theta z)^{m-1}$ (see Figure \ref{fig:RaPrdepth} (a)) and so the fluid becomes less supercritical at higher depths in the layer as $N_{\rho}$ is increased -- this we might expect to lead to slower velocities. (iii) The Prandtl number, $Pr$, is small at the top of the layer; low $Pr$ flows are known to favour flywheel convection where energy is transferred to kinetic energy via the dominance of the inertial terms. For small $Pr$ flows heat transport is known to be inefficient \citep{Jonesetal1976, CleverBusse1981, Ribeiroetal2015} e.g., \citet{CleverBusse1981} found convective heat transport to be relatively independent of $Pr$ for $Ra$ larger than a critical value and $0.001\leq Pr \leq 0.71$.

The corresponding kinetic energies for simulations in each of the regimes A-D are shown in Figure~\ref{fig:energies}(b). This measure is affected by the density of the fluid in addition to the fluid velocities and is defined as $KE=\langle\bar\rho (u^2+v^2+w^2)\rangle/2$. For the high Rossby number cases, $KE$ is a monotonically decreasing function of $N_{\rho}$; whilst the average velocity increases between $N_{\rho}=0$ and $N_{\rho}=1$, the decrease in the mean density is a more significant effect, leading to a smaller kinetic energy. By contrast, in the small Rossby number cases, the behaviour of $KE$ has a similar trend to that of $u_{rms}$.

%%%%%%%%%%%%%%%%%%%%%%%%%
%%%%%%%%%%%%%%%%%%%%%%%%%
%%%%%% NUSSELT NUMBER
%%%%%%%%%%%%%%%%%%%%%%%%%
%%%%%%%%%%%%%%%%%%%%%%%%%
To determine how the stratification affects the convective heat transport, we have calculated the Nusselt number (defined in (\ref{eqn:Nusselt})) for each of our simulations (see Figure~\ref{fig:energies}(c)). We see that for all cases here (weakly and rapidly rotating) increasing $N_{\rho}$ leads to less efficient heat transport (smaller $Nu$). This is consistent with the numerical and analytical results of \citet{MizerskiTobias2011}.
Since here we fix $Ra$ at the bottom of the domain, and the critical Rayleigh number is an increasing function of $N_{\rho}$, the supercriticality of the convection is decreased as $N_{\rho}$ is increased. This should correspond to a decrease in $Nu$ with $N_{\rho}$ and so in some sense the behaviour of $Nu$ in Figure~\ref{fig:energies}(c) is not a surprise. However, if the change in supercriticality were the only effect we would expect $KE$ and $u_{rms}$ to decrease monotonically with $N_{\rho}$ but, as discussed above, this is not the case. This suggests that the influence of stratification on heat transport is more complicated.

We find that $Nu$ for all the simulations examined here reaches a statistically-steady state well within a thermal diffusion time (even in regime (ii) where the kinetic energy may not reach a statistically-steady state for several diffusion times). This implies that the heat flux does not change much as the large-scale convective structures (investigated further in section \ref{subsec:LSflows}) continue to form over a slow time scale.

\begin{figure}
    \centering
    \includegraphics[scale=1]{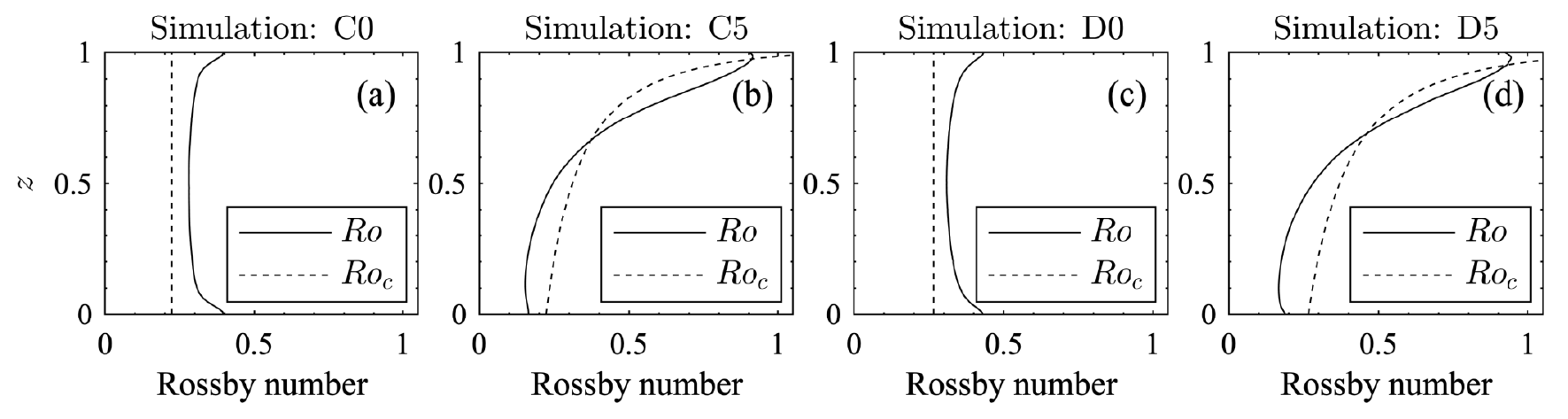}
    \caption{Rossby number as a function of depth for a case with $\phi=90^{\circ}$ and $N_{\rho}=0$ (a) (simulation C0), $N_{\rho}=5$ (b) (simulation C5) and for a case with $\phi=45^{\circ}$ and $N_{\rho}=0$ (c) (simulation D0), $N_{\rho}=5$ (d) (simulation D5)  In each case a Rossby number calculated from input parameters ($Ro_c$) and a Rossby number calculated from outputs of the simulation ($Ro$) is included.}
    \label{fig:Rossby}
\end{figure}

%%%%%%%%%%%%%%%%%%%%%%%%%
%%%%%%%%%%%%%%%%%%%%%%%%%
%%%%%% Rossby number input vs output here?
%%%%%%%%%%%%%%%%%%%%%%%%%
%%%%%%%%%%%%%%%%%%%%%%%%%
Figure \ref{fig:Rossby} shows $Ro$ (as defined in (\ref{eqn:Ro})) for rapidly rotating simulations with $N_{\rho}=0$ (Boussinesq) and $N_{\rho}=5$. In (a) and (b), $\phi=90^{\circ}$ and the rotation vector is aligned with gravity and in (c) and (d), $\phi=45^{\circ}$. In the Boussinesq cases, $Ro$ is roughly constant across the domain and is in the regime we expect rotation to dominate ($Ro\ll1$). For $N_{\rho}=5$, $Ro$ increases significantly from bottom to top of the domain suggesting a reduction in the influence of rotation as we move higher in the layer.
We have also plotted $Ro_c$ (which has the advantage it can be calculated before the simulation started) and it seems (at least for these cases) to perform well in describing the effect of rotation relative to inertia over the depth of the domain. 

%%%%%%%%%%%%%%%%%%%%%%%%%%%%%%%%%%%%%%%%%%%%%%%%%%%
%%%%%%%%%%%%%%%%%
%%%%%%%%%%%%%%%%% Large scale flows
%%%%%%%%%%%%%%%%%
%%%%%%%%%%%%%%%%%%%%%%%%%%%%%%%%%%%%%%%%%%%%%%%%%%%
\subsection{Large-scale flows}\label{subsec:LSflows}

\subsubsection{Regime (i): Weakly rotating}
\begin{figure}
    \centering
    \includegraphics[scale=1]{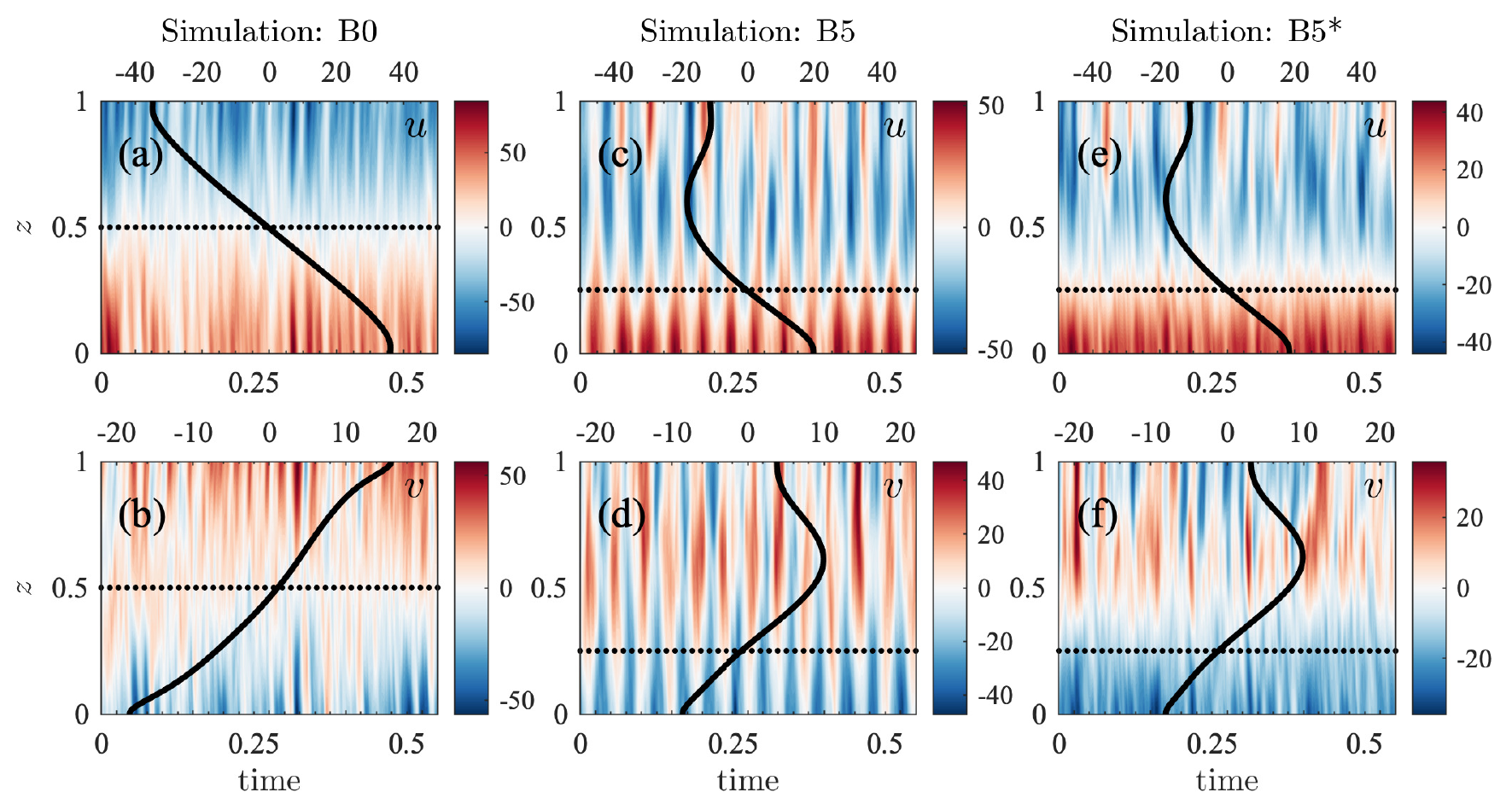}
    \caption{Horizontally averaged (mean) horizontal flow component $\langle u \rangle_{x,y}$ (top row), $\langle v \rangle_{x,y}$ (bottom row) as a function of time and $z$ (contours, colourbar) and time-averaged mean flow (black line, top axis) for simulations (a,b) B0, (c,d) B5 and (e,f) B5*. Note the only difference between simulations B5 and B5* is the state from which each simulation was initialised}. The horizontal dotted lines denote the depth at which the centre of mass lies.
    \label{fig:meanflow}
\end{figure}

As has been shown in previous studies (\citep{HS1983,curtob:2016} and references therein), convection is able to generate coherent mean (i.e., horizontally averaged) flows when the rotation vector is not aligned with gravity. For example, in Figure \ref{fig:meanflow} (a) and (b), systematic $\langle u \rangle_{x,y}$ and $\langle v \rangle_{x,y}$ are generated in time. In these cases, the system is Boussinesq and as expected there is a symmetry about $z=0.5$, as indicated by the dotted horizontal lines; above $z=0.5$, $u$ is predominantly negative and $v$ positive, whilst below $z=0.5$, $u$ is predominantly positive and $v$ negative. Keeping all other parameters fixed, and increasing $N_{\rho}$ to 5, leads to the mean flows shown in Figure \ref{fig:meanflow} (c) and (d). Whilst still present, the flows are now less systematic and more oscillatory; for example, $u$ oscillates between positive and negative flow across most of the layer depth but with a preference for negative flow in a time-averaged sense. At each time, the magnitude of the flows is similar to the Boussinesq case but the oscillations result in a smaller time-averaged flow.
We note that these $N_{\rho}=5$ flows were obtained from a simulation which started from an existing solution that has a non-zero vertically integrated horizontal momentum, and therefore the vertically integrated horizontal momentum is not necessarily zero at each time. This can be seen if we first consider the horizontally averaged, horizontal momentum equations:
\begin{equation}
    \frac{\partial }{\partial t}(\bar\rho \bar u) + \frac{\partial}{\partial z} ( \bar \rho \overline{uw})= Ta^{\frac{1}{2}} Pr \sin\phi \bar\rho \bar v + Pr \frac{\partial}{\partial z} \left(\bar \rho \frac{\partial \bar u}{\partial z}\right),
\end{equation}
\begin{equation}
    \frac{\partial }{\partial t}(\bar\rho \bar v) + \frac{\partial}{\partial z} ( \bar \rho \overline{vw})= -Ta^{\frac{1}{2}} Pr \sin\phi \bar\rho \bar u + Pr \frac{\partial}{\partial z} \left(\bar \rho \frac{\partial \bar v}{\partial z}\right).
\end{equation}
If we then integrate these over $z$, we find equations for the vertically integrated horizontal momentum:
\begin{equation}\label{eqn:vertinthorizmom1}
    \frac{\partial }{\partial t}\left(\int \bar\rho \bar u \,dz\right) = Ta^{\frac{1}{2}} Pr \sin\phi \int \bar\rho \bar v \,dz,
\end{equation}
\begin{equation}\label{eqn:vertinthorizmom2}
    \frac{\partial }{\partial t}\left(\int \bar\rho \bar v \,dz\right) = -Ta^{\frac{1}{2}} Pr \sin\phi \int \bar\rho \bar u \,dz 
\end{equation}
where we have made use of the velocity boundary conditions in (\ref{eq:velBCs}). 
It follows from (\ref{eqn:vertinthorizmom1}) and (\ref{eqn:vertinthorizmom2}) that if $\int \bar\rho \bar u \,dz$ and $\int \bar\rho \bar v \,dz$ are zero initially, they must remain so for all time, but not otherwise.
By differentiating (\ref{eqn:vertinthorizmom1}) and substituting into (\ref{eqn:vertinthorizmom2}) (and vice-versa) one can see that (when they are non-zero initially) the components of vertically integrated horizontal momentum are governed by wave equations.
Figure \ref{fig:meanflow} (e) and (f) show the equivalent case to (c) and (d) but starting from an initial condition with zero velocity. Unlike in (c) and (d) where there is significant transfer between $\bar u$ and $\bar v$ owing to the rotation, here there is no such transfer. The exchange between $\bar u$ and $\bar v$ in (c) and (d) results in a spiralling of the mean flow that is not observed in the case shown in (e) and (f) (see supplementary material for corresponding movies of simulations B5 and B5* which show most clearly the spiralling of the flow in simulation B5 but not in B5*).
We remark further that if we initialise a Boussinesq simulation with an initial condition that has a nonzero vertically integrated horizontal momentum (simulation B0*), then that case will also possess a non trivial vertically integrated horizontal momentum at later times (as governed by the wave equation resulting from equations (\ref{eqn:vertinthorizmom1}) and (\ref{eqn:vertinthorizmom2})).
However, even with a nonzero vertically integrated momentum, the mean flow in this Boussinesq case does not exhibit spiralling (see supplementary material for movies of the mean flows in simulations B0 and B0*). 
In summary, the stratification appears to introduce a symmetry breaking that leads to a spiralling mean flow (in those cases which initially have some vertically integrated horizontal momentum) that is not seen in the Boussinesq cases.
Note, we find bulk properties such as those displayed in Figure \ref{fig:energies} to be the same in both the two $N_{\rho}=5$ cases with different initial conditions and the two Boussinesq case with different initial conditions.

Regardless of the initial condition, stratification introduces an asymmetry in the layer; the depth in the layer at which $\langle u \rangle_{x,y}$ and $\langle v \rangle_{x,y}$ cross zero occurs approximately around the depth at which the centre of mass lies (this is marked with the horizontal dotted line in the plots). This transition depth was also suggested in the fully compressible studies of \citet{Brummelletal1998} and is a consequence of momentum conservation and stress-free boundary conditions.

The self-consistently generated mean flow that is produced in this regime has strong vertical shear which may have important consequences for dynamos and the viability of such flows to drive a large-scale dynamo are discussed further in section \ref{subsec:helicity}.

\subsubsection{Regime (ii): Rapidly rotating}
\begin{figure}
    \centering
    \includegraphics[scale=1]{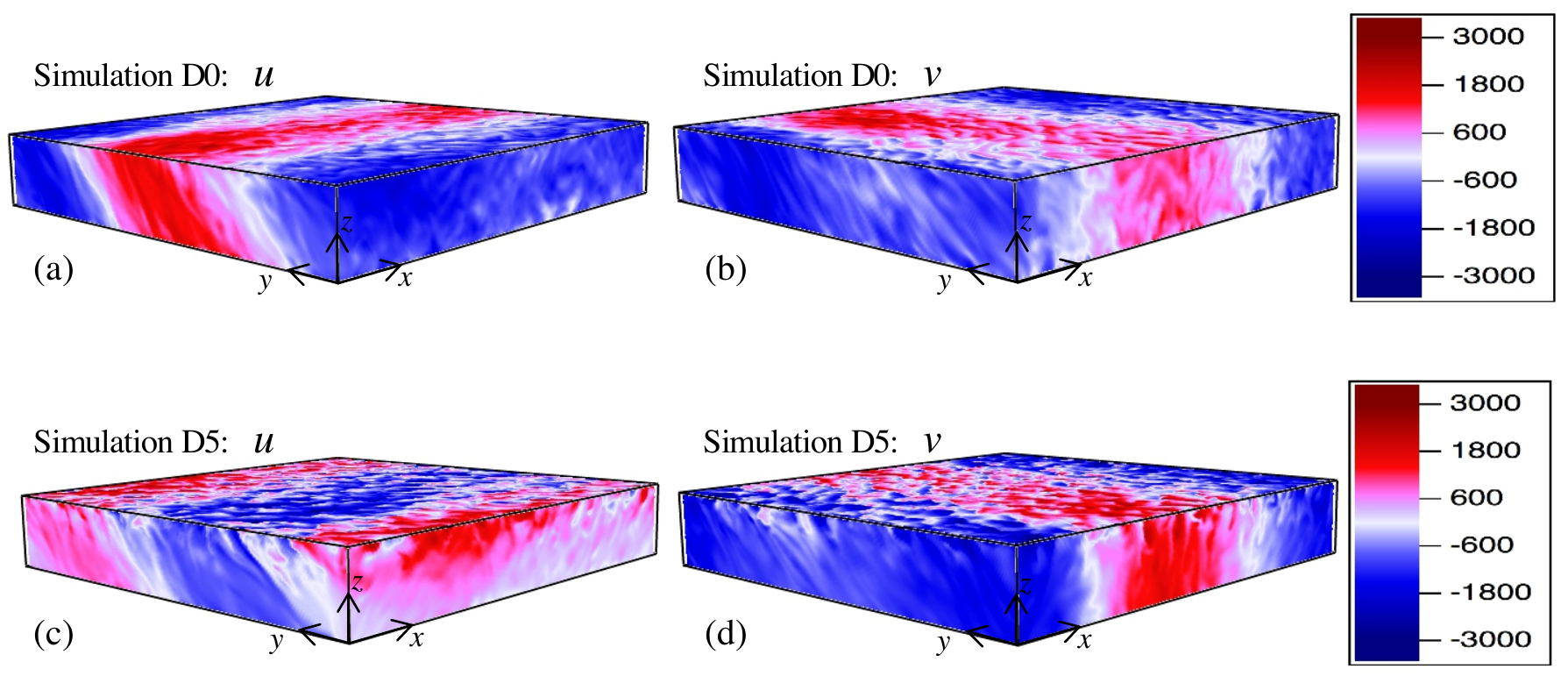}
    \caption{Snapshots of the horizontal velocities, $u$ (left column) and $v$ (right column) for two different simulations. (a,b): simulation D0 ($N_{\rho}=0$) and (c,d): simulation D5 ($N_{\rho}=5$).}
    \label{fig:uvsnapshots}
\end{figure}

In regime (ii) (simulation subsets C and D), $Ro_c$ is much smaller than in regime (i). In these cases there are no systematic mean flows like those discussed in the previous section but instead large-scale coherent jets form which often correspond to a large-scale vortex.
These structures have been investigated in a number of different systems; in compressible convection by \citet{Mantereetal2011} and \citet{ChanMayr2013}, in an asymptotic model with polar rotation by \citet{Rubioetal2014} and in a Boussinesq model by \citet{Favieretal2014} and \citet{Guervillyetal2014}. More recently, they were observed at an arbitrary latitude by \citet{Novietal2019} and \citet{Currieetal2020}.
A classic example of these flows is given in Figure \ref{fig:uvsnapshots} where the left (right) hand column shows $u$ ($v$) for a Boussinesq case ((a) and (b)) and a case with $N_{\rho}=5$ ((c) and (d)). In both cases, $u$ ($v$) varies very little along the $x$ ($y$) direction and contains one wavelength in $y$ ($x$); it is clear that averaging over $y$ ($x$) will lead to the positive and negative jets cancelling out and therefore no systematic mean flows exists in these cases.

\begin{figure}
    \centering
    \includegraphics[scale=1]{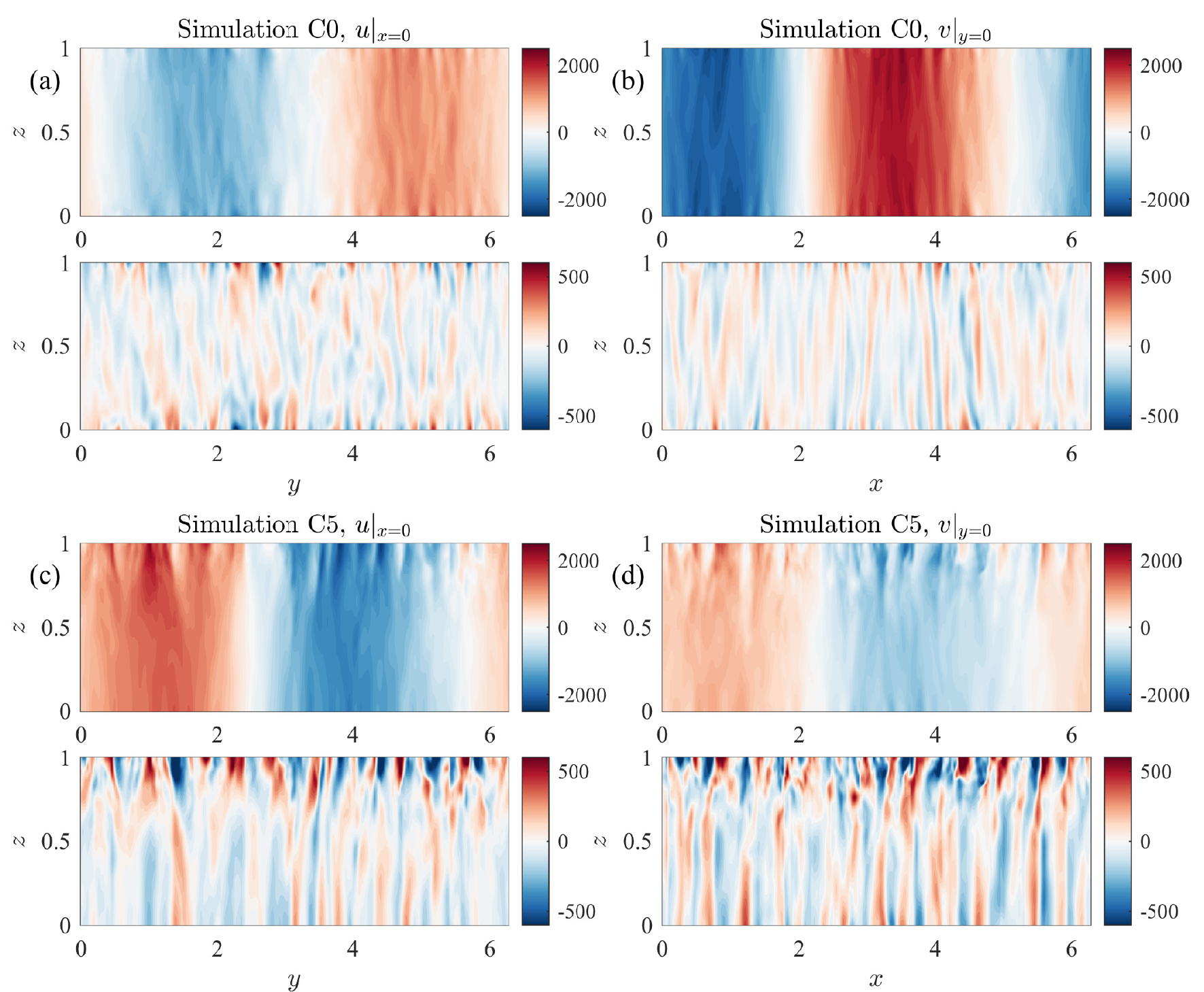}
    \caption{Slices of the horizontal velocities. For $u$ (left column), the slices are taken at $x=0$ and for $v$ (right column), the slices are taken at $y=0$. In (a) and (b), $N_{\rho}=0$ (simulation C0) and in (c) and (d), $N_{\rho}=5$ (simulation C5). Within each subfigure, the top panel gives the slice of the velocity component in that plane, and the bottom panel gives the perturbations that remain when the average along the rotation axis is subtracted.}
    \label{fig:TP}
\end{figure}

Figure \ref{fig:snapshots} suggested that whilst in the Boussinesq case the convective cell structures are largely independent of the direction of the rotation axis, at strong stratification there is variation along $\Omega$. This effect can also be seen to some extent in the stratified snapshots of $u$ and $v$ in Figure \ref{fig:uvsnapshots} (c) and (d) respectively.
This reduction in the extent to which the Taylor-Proudman constraint is satisfied as stratification is increased is exhibited more clearly in Figure \ref{fig:TP}, where vertical slices of the flow and the corresponding perturbations that remain after an average along the rotation axis has been subtracted, are shown.
In the Boussinesq cases (a) and (b), the departures from a perfect Taylor-Proudman state are fairly uniform across the depth of the layer. By contrast, for $N_{\rho}=5$, the departures are much larger at the top of the domain. In addition, the magnitude of the perturbations are relatively larger in the strongly stratified case. 

Mathematically speaking, the departures from Taylor-Proudman can be seen by the extra terms present in the vorticity equation. The $x$, $y$ and $\Omega$ (i.e., along the direction of the rotation axis) components of the vorticity equation (formed by taking the curl of equation (\ref{newmom})) are as follows:
\begin{equation}\label{eqn:vortx}
    \frac{\partial \omega_x}{\partial t} + (\bm{u}\cdot\nabla)\omega_x - (\bm{\omega}\cdot\nabla)u - \frac{m\theta w}{1+\theta z}\omega_x = Ta^{\frac{1}{2}}Pr(\Omega\cdot\nabla)u + RaPr\frac{\partial S}{\partial y} + Pr\nabla^2\omega_x + \frac{Pr m \theta}{1+\theta z}\left(\frac{\partial \omega_x}{\partial z} + \frac{(1+2m)\theta}{3(1+\theta z)}\frac{\partial w}{\partial y}\right),
\end{equation}

\begin{align}\label{eqn:vorty}
    \frac{\partial \omega_y}{\partial t} + (\bm{u}\cdot\nabla)\omega_y - (\bm{\omega}\cdot\nabla)v - \frac{m\theta w}{1+\theta z}\omega_y& = Ta^{\frac{1}{2}}Pr\cos\phi\frac{m\theta w}{1+\theta z} + Ta^{\frac{1}{2}}Pr(\Omega\cdot\nabla)v - RaPr\frac{\partial s}{\partial x} \nonumber\\
    &+ Pr\nabla^2\omega_y + \frac{Pr m \theta}{1+\theta z}\left(\frac{\partial \omega_y}{\partial z} - \frac{(1+2m)\theta}{3(1+\theta z)}\frac{\partial w}{\partial x}\right),
\end{align}

\begin{align}\label{eqn:vortOm}
    \frac{\partial \omega_\Omega}{\partial t} + (\bm{u}\cdot\nabla)\omega_\Omega - (\bm{\omega}\cdot\nabla)u_\Omega - \frac{m\theta w}{1+\theta z}\omega_\Omega &= Ta^{\frac{1}{2}}Pr\frac{m\theta w}{1+\theta z}+ Ta^{\frac{1}{2}}Pr(\Omega\cdot\nabla)u_\Omega - RaPr\cos\phi\frac{\partial s}{\partial x}  \nonumber\\
    &+Pr\nabla^2\omega_\Omega + \frac{Pr m \theta}{1+\theta z}\left(\frac{\partial \omega_\Omega}{\partial z} - \frac{(1+2m)\theta}{3(1+\theta z)}\cos\phi\frac{\partial w}{\partial x}\right),
\end{align}
where $\bm{\omega}=\nabla \times \bm{u} = (\omega_x,\omega_y,\omega_z)$ is the vorticity and $\omega_{\Omega}=\bm\omega\cdot\bm \Omega$, $u_{\Omega}=\bm u\cdot\bm \Omega$ are the components of $\bm\omega$ and $\bf u$ along $\bm\Omega$, respectively.

If $Ro$ is small, we expect $Ta^{\frac{1}{2}}Pr(\bm{\Omega}\cdot\nabla)u \sim0$ to be the dominant balance in (\ref{eqn:vortx}) and so variations in $u$ should be independent along $\bm{\Omega}$ irrespective of $N_{\rho}$. In (\ref{eqn:vorty}) a departure from $Ta^{\frac{1}{2}}Pr(\bm{\Omega}\cdot\nabla)v \sim0$ is introduced if $\phi\neq\frac{\pi}{2}$ and $\theta\neq0$ (and is given by $Ta^{\frac{1}{2}}Pr\cos\phi\frac{m\theta w}{1+\theta z}$). Similarly, in (\ref{eqn:vortOm}), a departure from $Ta^{\frac{1}{2}}Pr(\bm{\Omega}\cdot\nabla)u_{\Omega} \sim0$ is introduced if $\theta\neq0$ (given by $Ta^{\frac{1}{2}}Pr\frac{m\theta w}{1+\theta z}$). This suggests that departures from a perfect Taylor-Proudman state in $u$ are a result of the nonlinear, inertial terms contributing significantly. In fact, if we consider the depth dependent $Ro$ for the cases shown in Figure \ref{fig:TP} (see Figure \ref{fig:Rossby} (a) and (b)) then $Ro$ is order 1 at the top of the domain which suggests the inertia terms are not completely dominated by the Coriolis force. This effect may also be present in causing departures from Taylor-Proudman in $v$ and $w$. Moreover, we would expect that in a regime where $Ro$ is small across the whole depth, departures from Taylor-Proudman can still be introduced by the additional terms in (\ref{eqn:vorty}) and (\ref{eqn:vortOm}) introduced by stratification.

\begin{figure}
    \centering
    \includegraphics[scale=0.9]{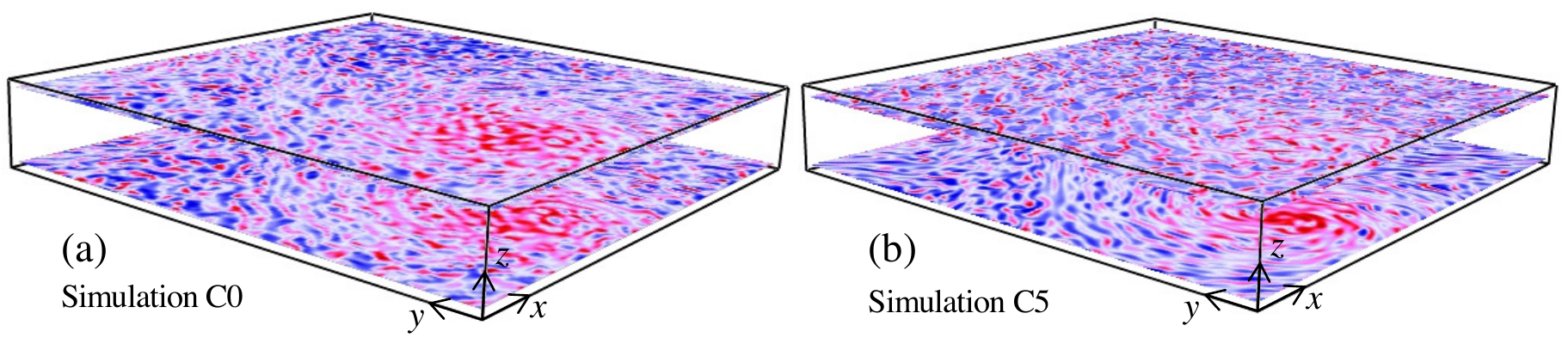}
    \caption{Snapshots of the vertical vorticity for (a) simulation C0 ($N_{\rho}=0$) and (b) simulation C5 ($N_{\rho}=5$). In (a) a large-scale vortex is clearly seen to extend across the depth of the domain, whilst in (b) the vortex is only seen at the bottom of the domain.}
    \label{fig:VortexRossby}
\end{figure}

As mentioned before, the large-scale structures in the velocity components may correspond to a large-scale vortex as seen in several previous studies.
To examine these structures, the component of vorticity along the rotation axis is shown in Figure \ref{fig:VortexRossby}. In the Boussinesq case, a concentrated region of positive vorticity can be seen in the layers near the top and bottom of the layer. This is indicative of a vortex that extends across the whole depth in line with what has been seen in previous studies. However, in the strongly stratified case, the vortex can clearly be seen near the bottom of the layer but not near the top of the domain. 
We note that the vortex appears to maintain the same sign of vorticity as the the stratification is changed.

The prevalence of an inverse cascade (leading to a vortex) can be examined further by considering a shell-to-shell energy transfer analysis \citep[][]{Alexakisetal2005,Favieretal2014}. We focus on the transfer function $T(Q,K,z)$ which represents the energy transfer from shell Q to shell K at a height $z$. In this context, a shell K is defined in wavenumber space as those wavenumbers that satisfy $K<k_h\leq K+1$, where $k_h=\sqrt{k_x^2+k_y^2}$ is the horizontal wavenumber ($k_x$ and $k_y$ are the wavenumbers in the $x$ and $y$ directions respectively). We then define a filtered field $\bm{u}_K$ such that
\begin{equation}
  \bm{u}_K(x,y,z)=\sum_{K<k_h\leq K+1}\bm{\hat u}(k_x,k_y,z)e^{ik_xx+ik_yy},
\end{equation}
and then, $T(Q,K,z)$ is given (for all $z$) by
\begin{equation}
    T(Q,K,z)=-\int \bar \rho \bm{u}_K\cdot[(\bm{u}\cdot\nabla)\bm{u}_Q]\,dxdy.
\end{equation}
$T(Q,K,z)$ can be interpreted as follows: if $T(Q,K,z)$ is positive, then a positive amount of energy is extracted from shell Q and given to shell K. 
Note, if an average is taken over all depths $z$ then $\tilde T(Q,K)=-\tilde T(K,Q)$ where $\tilde T=\int T\, dz$ but this does not have to hold at each $z$.

\begin{figure}
    \centering
    \includegraphics[scale=1]{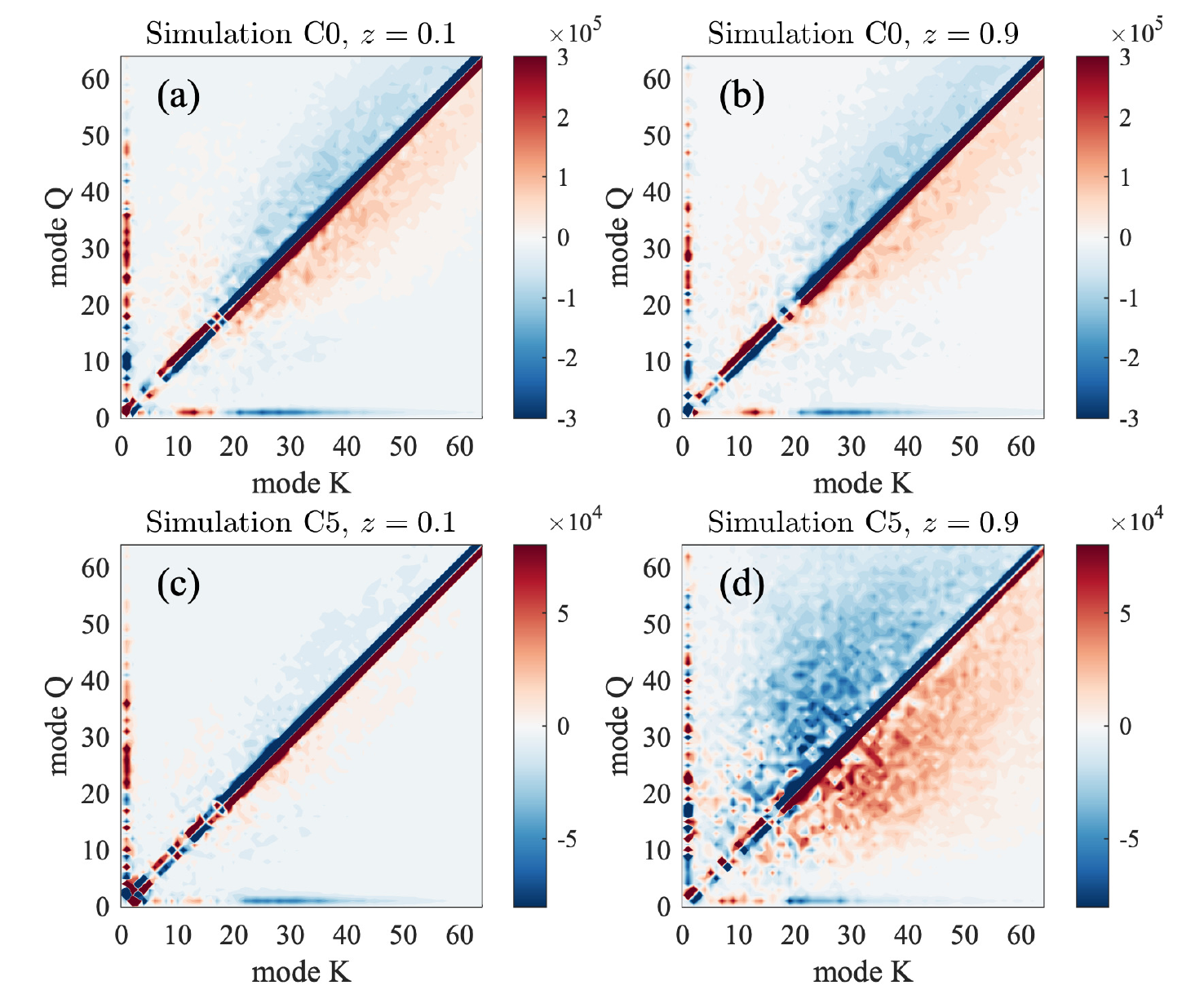}
    \caption{Slices at a fixed depth of time-averaged transfer functions $T(Q,K,z)$ for shells with $0<Q,K<64$. (a) and (b) show slices at $z=0.1$ and $z=0.9$, respectively, for simulation C0 ($N_{\rho}=0$) and (c) and (d) show slices at $z=0.1$ and $z=0.9$, respectively, for simulation C5 ($N_{\rho}=5$). The function is positive when energy is extracted from shell Q and given to shell K.}
    \label{fig:TransferFunctions}
\end{figure}

In Figure \ref{fig:TransferFunctions}, we plot $T(Q,K,z=0.1)$ and $T(Q,K,z=0.9)$ averaged over several snapshots for $N_{\rho}=0$ ((a) and (b)) and for $N_{\rho}=5$ ((c) and (d)). Here we see that, for $N_{\rho}=0$ the results are entirely consistent with those in \citet{Favieretal2014}; the large-scale structure is fed by non-local energy transfer into large-scale modes. As the structure is baroclinic, the transfer is the same at $z=0.1$ and $z=0.9$. In contrast, for the $N_{\rho}=5$ stratified case, whilst the non-local transfer is still effective towards the bottom of the domain (in the low $Ro$ region), it is significantly disrupted at $z=0.9$. Hence relatively less energy is channeled to the coherent structure.

As with the reduction in the Taylor-Proudman constraint that occurs near the top of the domain, the non-existence of a vortex there can perhaps be explained by considering the Rossby number as a function of depth. $Ro$ was shown for these cases in Figure \ref{fig:Rossby} (a) and (b) and as discussed there it is much larger at the top of the domain than at the bottom, when $N_{\rho}=5$. This implies the rotational dominance is much reduced at the top of the domain.

%%%%%%%%%%%%%%%%%%%%%
%%%%%%%%%%%%%%%%%%%%%
%%%%%  Helicity
%%%%%%%%%%%%%%%%%%%%%
%%%%%%%%%%%%%%%%%%%%%
\subsection{Effect of stratification on net helicity}\label{subsec:helicity}
From a dynamo perspective, we are interested in if these flows can generate a large-scale, systematic, magnetic field. In this case, lack of reflectional symmetry often as manifested by kinetic helicity is believed to be an important quantity \cite{MoffattDormy2019}.
The relative helicity is given by
\begin{equation}\label{relhel}
h(z)\,=\left\langle\frac{\langle \bm u' {\bm \cdot} \bm \omega' \rangle_{x,y}}{\langle \bm u'^2\rangle_{x,y}^{{1}/{2}}\langle \bm \omega'^2\rangle_{x,y}^{{1}/{2}}}\right\rangle_{\!\!t}\,,
\end{equation}
where $\bm \omega'=\bm\nabla\times\bm u'$ is the vorticity, and $\langle{\cdot}\rangle_t$ denotes an average over $t$. $\bm u'=\bm u-\langle\bm u\rangle_{x,y}$ is the fluctuation of $\bm u$ about its mean state $\langle\bm u\rangle_{x,y}$. Since we are really interested in the helicity of the turbulent eddies and not the large-scale component of the flow, we use $\bm u'$ in the calculation of $h$.

\begin{figure}
    \centering
    \includegraphics[scale=1]{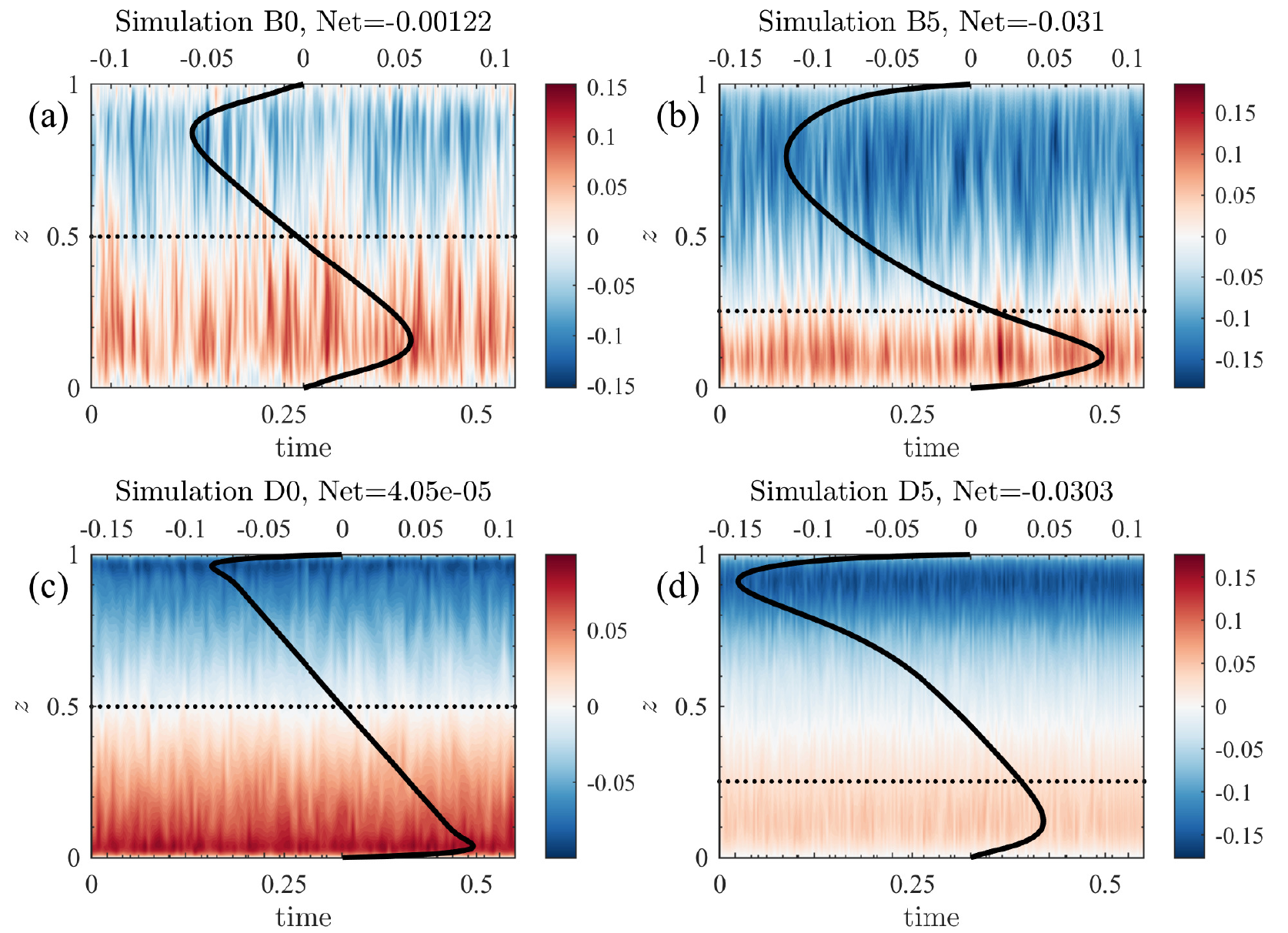}
    \caption{Relative helicity in the perturbations (i.e., after the horizontal mean flow has been removed) as a function of $z$ and time (contours, colorbar) for simulation B0 (a), simulation B5 (b), simulation D0 (c) and simulation D5 (d). The time-averaged relative helicity is given by the over-plotted black lines (top axes). The net helicity (time and vertically averaged value) for each case is displayed in the title for each panel. The horizontal dotted lines denote the depth at which the centre of mass lies. }
    \label{fig:helicity}
\end{figure}

In the Boussinesq cases, we expect an average of $h(z)$ over $z$ to lead to zero net helicity and this is what is essentially seen in Figure \ref{fig:helicity} (a) and (c) where the negative helicity in the top half of the layer cancels with the positive helicity in the bottom half.
As we have seen, stratification introduces an asymmetry; this results in the flows possessing a non-zero net helicity, when averaged over the whole domain (see Figure \ref{fig:helicity} (b) and (d)). It has been shown by \citet{CattaneoTobias2014}, \citet{Nigroetal2017} and \citet{Pongkitiwanichakuletal2016} that large-scale magnetic field can be generated even for turbulent flows if the product of the shear and helicity is large enough.  
However, for the cases studied here, even though we find the net helicity increases approximately linearly with $N_{\rho}$, it is small even for $N_{\rho}=5$ and it remains to be seen if this is enough to generate large-scale dynamo waves. This might depend on the location of strongest shear.

%%%%%%%%%%%%%%%%%%%%%
%%%%%%%%%%%%%%%%%%%%%
%%%%%  Conclusions
%%%%%%%%%%%%%%%%%%%%%
%%%%%%%%%%%%%%%%%%%%%
\section{Discussion \& Conclusions}

In this paper, we have examined the effect of stratification on two key regimes of rotating anelastic convection, with both a vertical and tilted (to the direction of gravity) rotation vector. These regimes correspond to  (i) weakly rotating ($Ro>1$) and (ii) rapidly rotating ($Ro \ll 1$) flows. It is known that density stratification introduces an asymmetry that leads to significantly faster velocities at the top of the domain. Moreover, the variation in $Ra$ and  $Pr$ in strongly stratified cases for the setup considered here means the convection may behave quite differently at different depths, thus having a significant effect on heat transport.

For both regimes, strong systematic, large-scale flows can be driven but they vary significantly in their character depending on the rotational constraint. In regime (i), if the rotation vector is tilted from the vertical, then a systematic mean flow (with vertical shear) is driven; here the mean is taken over horizontal planes. Stratification has the effect of ``asymmetrising" the mean flow so that it is no longer symmetric about the mid depth. 
The addition of stratification can (in cases that start out with a non-zero vertically integrated horizontal momentum) also introduce a symmetry breaking that leads to a net spiralling of the mean flow, not seen in the Boussinesq system (even when the system is initialised with a state that has a non-zero vertically integrated horizontal momentum). In regime (ii), the systematic flows take the form of coherent structures, that can correspond to a large-scale vortex. For a fixed Rossby number at the bottom of the domain, the stratification has the effect of increasing $Ro$ (decreasing the rotational constraint) as one moves upwards in the layer. This coincides with larger (than in the Boussinesq case) departures from a Taylor-Proudman state and also the loss of a coherent vortex near the top of the domain, owing to modification of the energy transfer mechanisms.

Despite the simplicity of the model considered here, some of the underlying physics e.g., the role of stratification in modifying convection, may still be relevant in many astrophysical contexts.
It is interesting to speculate on the implications of our results for large-scale structures in giant planets.  We believe that the effect of stratification here is to disrupt large-scale vortex structures at rapid rotation, owing to changes in the Rossby number across the convective structure. However, it is unclear what the coherence of the vortex would be in (more rapidly-rotating) cases where the Rossby number remains small for the whole structure. Hence we are wary of overstating the implication of these calculations for the vortices found in Jupiter and Saturn at this stage.

We also note that shear is expected to play an important role in the generation of magnetic field through dynamo action. In addition, helicity is thought to be important for the generation of large-scale magnetic field. For both regimes considered here, there is significant helicity in the layer and the key difference introduced by stratification is that the flows now possess net helicity. However, the net helicity for the cases considered here is still relatively small and so it remains to be seen if this helps in the generation of dynamo waves at high magnetic Reynolds number, $Rm$. This suggests that the next step should involve the investigation of the dynamo properties of flows such as those examined here --- and this work is currently underway.

%%%%%%%%%%%%%%%%%%%%%
%%%%%%%%%%%%%%%%%%%%%
%%%%%  Acknowledgments
%%%%%%%%%%%%%%%%%%%%%
%%%%%%%%%%%%%%%%%%%%%
\begin{acknowledgments}
The authors would like to acknowledge two anonymous referees for helpful and constructive reviews of the initial manuscript.
LKC acknowledges support from STFC Grant ST/R000891/1 and the European Research Council under ERC grant agreement No.~337705 (CHASM).
SMT would like to acknowledge support of funding from the European Research Council (ERC) under the European Union Horizon 2020 research and innovation programme (grant agreement no.~D5S-DLV-786780).
The calculations for this paper were performed on the University of Exeter Supercomputer, Isca, part of the University of Exeter High-Performance Computing (HPC) facility. Additional simulations were performed using the DiRAC Data Intensive service at Leicester, operated by the University of Leicester IT Services, which forms part of the STFC DiRAC HPC Facility (www.dirac.ac.uk). This equipment was funded by BEIS capital funding via STFC capital grants ST/K000373/1 and ST/R002363/1 and STFC DiRAC Operations grant ST/R001014/1. DiRAC is part of the National e-Infrastructure.
The 3D renderings in figures \ref{fig:snapshots}, \ref{fig:uvsnapshots} and \ref{fig:VortexRossby} were created using VAPOR \citep{vapor}.

\end{acknowledgments}

%%%%%%%%%%%%%%%%%%%%%
%%%%%%%%%%%%%%%%%%%%%
%%%%%  Appendix
%%%%%%%%%%%%%%%%%%%%%
%%%%%%%%%%%%%%%%%%%%%

%\nocite{*}

\bibliography{references_all}% Produces the bibliography via BibTeX.

\end{document}